\documentclass[reqno]{conm-p-l}

\usepackage{amssymb, amscd,enumerate,enumitem,color, hyperref,
mathtools, relsize, bm, graphicx}
\usepackage[british]{babel}
\usepackage[mathscr]{euscript}

\newtheorem{theorem}{Theorem}[section]

\newtheorem{theo}{Theorem}[section]
\newtheorem{lem}[theo]{Lemma}
\newtheorem{prop}[theo]{Proposition}

\newtheorem{cor}[theo]{Corollary}

\theoremstyle{definition}
\newtheorem{definition}[theorem]{Definition}

\theoremstyle{remark}

\numberwithin{equation}{section}

\newenvironment{pf}{\noindent{\it Proof. }}{$\square$\par\medskip}

\newcommand{\beq}{\begin{equation}}
\newcommand{\eeq}{\end{equation}}
\renewcommand{\a}{\alpha}

\newcommand{\f}{\varphi}

\renewcommand{\r}{\rho}
\newcommand{\s}{\sigma}

\newcommand{\bC}{\mathbb{C}}

\newcommand{\bD}{\mathbb{D}}
\newcommand{\bR}{\mathbb{R}}
\newcommand{\bZ}{\mathbb{Z}}

\newcommand{\bH}{\mathbb{H}}
\newcommand{\bK}{\mathbb{K}}

\newcommand{\bO}{\mathbb{O}}

\newcommand{\bT}{\mathbb{T}}

\renewcommand{\gg}{\mathfrak{g}}

\newcommand\GL{\mathrm{GL}}

\newcommand{\cA}{\mathscr{A}}
\newcommand{\cB}{\mathscr{B}}
\newcommand{\cC}{\mathcal{C}}
\newcommand{\cD}{\mathscr{D}}

\newcommand{\cH}{\mathscr{H}}

\newcommand{\cN}{\mathscr{N}}

\newcommand{\cS}{\mathscr{S}}
\newcommand{\cT}{\mathscr{T}}
\newcommand{\cU}{\mathscr{U}}
\newcommand{\cV}{\mathscr{V}}

\renewcommand{\square}{\kern1pt\vbox
{\hrule height 0.6pt\hbox{\vrule width 0.6pt\hskip 3pt
\vbox{\vskip 6pt}\hskip 3pt\vrule width 0.6pt}\hrule height0.6pt}\kern1pt}

\DeclareMathOperator\Aut{Aut}

\newcommand\Hom{\operatorname{Hom}}

\renewcommand\={:=}

\newcommand{\bt}{\begin{theo}\ \ }
\newcommand{\et}{\end{theo}}
\newcommand{\bp}{\begin{prop}\ \ }
\newcommand{\ep}{\end{prop}}
\newcommand{\bc}{\begin{cor}\ \ }
\newcommand{\ec}{\end{cor}}
\newcommand{\bl}{\begin{lem}\ \ }
\newcommand{\el}{\end{lem}}
\newcommand{\bd}{\begin{definition}}
\newcommand{\ed}{\end{definition}}
\newcommand{\be}{\begin{equation}}
\newcommand{\ee}{\end{equation}}

\def\<#1,#2>{\langle\,#1,\,#2\,\rangle}

\newcommand{\arr}{\begin{array}{rlll}}
\newcommand{\ea}{\end{array}}
\newcommand{\bea}{\begin{eqnarray}}
\newcommand{\eea}{\end{eqnarray}}
\newcommand{\bean}{\begin{eqnarray*}}
\newcommand{\eean}{\end{eqnarray*}}

\newcommand{\ve}{\varepsilon}

\newcommand{\bdot}{{\boldsymbol{\cdot}}}

\newcommand{\Nil}{\operatorname{Nil}}
\newcommand{\Solv}{\operatorname{Solv}}
\newcommand{\Herm}{\operatorname{\it Herm}}
\newcommand{\Hess}{\operatorname{Hess}}

\begin{document}

\title[Special Vinberg cones  and  invariant  admissible cubics]{Special Vinberg cones, invariant admissible cubics \\ and special real manifolds}


\author[D. V. Alekseevsky]{Dmitri V. Alekseevsky}
\address{Institute for Information Transmission Problems,
B. Karetnuj per. 19
Moscow 127051, Russia\ \&
    University of Hradec Králové,
    Faculty of Science,
    Rokitanského 62
   500 03 Hradec Králové, Czech Republic}
\curraddr{}
\email{dalekseevsky@iitp.ru}
\thanks{}

\author[A. Marrani]{Alessio Marrani}
\address{Instituto de F\'\i sica Teorica,
Universidad de Murcia,
Campus de Espinardo,
E-30100, Murcia, Spain}
\curraddr{}
\email{alessio.marrani@um.es}
\thanks{}

\author[A. Spiro]{Andrea Spiro*}
\address{Scuola di Scienze e Tecnologie,
Universit\`a di Camerino,
Via Madonna delle Carceri,
I-62032 Camerino (Macerata),
Italy}
\curraddr{}
\email{andrea.spiro@unicam.it}
\thanks{*Corresponding author}

\date{}

\begin{abstract} By Vinberg theory any homogeneous convex  cone $\cV$ may be realised   as the cone of  positive Hermitian
matrices in  a $T$-algebra of generalised matrices. The level hypersurfaces $\cV_{q}  \subset \cV$  of  homogeneous cubic polynomials $q$  with  positive definite Hessian (symmetric) form $g(q) \= - \Hess(\log(q))|_{T \cV_q}$ are   the  {\it  special real manifolds}.
Such manifolds occur as  scalar manifolds of the vector multiplets in $N=2$, $D=5$ supergravity and, through the $r$-map,  correspond to K\"ahler  scalar manifolds in $N = 2$ $D = 4$ supergravity.  We  offer a simplified exposition of   the Vinberg theory  in terms of   $\Nil$-algebras (= the subalgebras of upper triangular matrices in Vinberg  $T$-algebras) and  we  use it to describe  all  rational functions on a special Vinberg cone  that are $G_0$- or $G'$- invariant, where  $G_0$ is the  unimodular subgroup of
the solvable group $G$   acting simply transitively on the cone,  and $G'$ is the unipotent radical  of $G_0$. The results are used to determine   $G_0$- and $G'$-invariant  cubic polynomials $q$ that are {\it admissible} (i.e. such that the  hypersurface  $  \cV_q=\{ q=1\}\cap \cV $  has    positive definite Hessian form $g(q)$)   for  rank $2$ and rank $3$  special Vinberg cones. We get in this way examples of continuous  families of non-homogeneous special real  manifolds of cohomogeneity less than or equal to two.
\end{abstract}

\maketitle

\setcounter{section}0
\def\sideremark#1{\ifvmode\leavevmode\fi\vadjust{
\vbox to0pt{\hbox to 0pt{\hskip\hsize\hskip1em
\vbox{\hsize3cm\tiny\raggedright\pretolerance10000
\noindent #1\hfill}\hss}\vbox to8pt{\vfil}\vss}}}
\section{Introduction}
 In \cite{Vi} Vinberg  proved that any  homogeneous convex cone may be realised  as   the  cone  $\cV$ of positive definite matrices  in an   appropriate  algebra $\cH=Herm_m$ of generalised Hermitian matrices of rank $m$.
In \cite{AC1},  Alekseevsky and Cort\'es described  a  class  of  rank 3 homogeneous convex cones, called {\it special  Vinberg cones}.
Such a class of cones is associated to $%
\mathbb{Z}_{2}$-graded Clifford modules with invariant metrics. If the metric
is indefinite, Vinberg's construction produces homogeneous not
necessarily convex cones, named {\it indefinite} special Vinberg cones.
Alekseevsky and Cort\'es also proved  that   the  associated  {\it determinant   hypersurface} $\cV_d = \{d = 1\}\subset  \cV  $, given by  the  so-called cubic determinant $d$,   is  a   homogeneous special real manifold in    the sense of de Wit  and Van Proeyen. We recall that a   {\it special real manifold}    is   a hypersurface    $ S\subset  \bR^n$,  which is a subset of  the level set   $  \{  q=1\}$ of a  homogeneous   cubic polynomial $q$, and such that  the restriction  to $S$  of the  Hessian (symmetric bilinear)    form $g(q) \= - \Hess \log q$, defined  on  a neighbourhood of  $S$,  is     a Riemannian metric  on $S$  (see \cite{Cortes-red}). If this holds, the  cubic $q$ is called  {\it admissible}.\par
\smallskip
The  special real manifolds occur  in Physics  as  target spaces -- usually called {\it scalar manifolds} -- for the maps that represent  the scalar fields in the vector multiplets in $N = 2$, $D = 5$ supergravity.   Furthermore, by dimensional reduction,   any   real special scalar manifold  $\cV_d$  corresponds to a $2n$-dimensional K\"ahler manifold $\cS$, called {\it very special K\"ahler manifold}, which is a   target space for the scalar fields in the vector  multiplets in $N = 2$, $D = 4$ supergravity. This correspondence   is the
 {\it supergravity $r$-map} defined by De Wit and Van Proeyen and   is a special case of  the  Piatetski-Shapiro construction that   associates with any real convex cone $\cV \subset \bR^n$  a complex Siegel domain of the first kind
$\cS =  \bR^n  + i \mathcal{V}\subset  \bR^n + i \bR^n \subset \bC^n$. In turn, again by dimensional reduction, any (not necessarily very) special K%
\"{a}hler manifold corresponds to a $(4n+4)$-dimensional quaternionic
manifold, which is the target space for the scalar fields (after full
dualization of the 1-forms) in $N=4$, $D=3$ supergravity, and also the target
space for the scalar fields of hypermultiplets coupled to $N$-extended
supergravity in $D= 6$ $(N = (1,0))$, $5,4$ ($N=2$), and 3 ($N=4$) \cite{BW}. This
correspondence is the {\it supergravity c-map} \cite{CFG}. As mentioned, the real, K\"ahler and quaternionic geometries related by dimensional reduction (or, equivalently, by the r-map and c-map) are named {\it special} geometries; see e.g. \cite{WVP,LVP} for a comprehensive treatment and a list of references.\par
\smallskip
\par
Vinberg \cite{Vi} developed a theory of homogeneous convex cones  and 
gave a description of them as cones of positive definite Hermitian matrices in generalised matrix
algebras, named {\it Vinberg T-algebras}. This
class of homogeneous spaces has important applications in  several areas, as for instance  supergravities with eight supersymmetries in dimensions $3 \leq D \leq 6$ as well as  in the description of the entropy of their extremal black hole solutions  and in  geometric approaches to information
theory (for the latter, see e.g.  \cite{Combe-Manin,Mar,Bar}).\par
\smallskip
Alekseevsky \cite{Alek1} gave the classification of
homogeneous quaternionic K\"{a}hler manifolds,  which admit a solvable
transitive Lie group of isometries, and associated   an homogeneous K%
\"{a}hler manifold of half dimension to any  such manifold, with the exception of the  hyperbolic quaternionic spaces. Cecotti \cite{Cec} remarked that
such K\"{a}hler manifolds are exactly the homogeneous target spaces of
scalar fields of vector multiplets coupled to $N=2$, $D=4$ supergravity. In \cite{WP1} de Wit and Van Proeyen
classified the admissible cubics corresponding to homogeneous special real
manifolds. By the supergravity r-map, this in turn yields the classification
of homogeneous special K\"{a}hler manifolds with holomorphic prepotentials
given by  the aforementioned admissible cubics. A few years later,
Cort\'es \cite{Co} gave another proof of this classification, using the theory
of homogeneous K\"{a}hler manifolds of Piatetski-Shapiro, Gindikin and
Vinberg. The classification of such manifolds, determined   in \cite{WP1} and 
\cite{Co},  shows that they are exactly the homogeneous special K\"{a}hler
manifolds associated to the quaternionic K\"{a}hler manifolds classified in \cite%
{Alek1}. It is here worth remarking that Alekseevsky conjectured that any
homogeneous quaternionic K\"{a}hler manifold with negative scalar curvature
should be  one of those classified  in \cite{Alek1}. 
This conjecture has been recently proved to be true by B\"{o}hm and Lafuente in  \cite{BL}.\par
\smallskip
The homogeneous  special real manifolds, that are  determined by  {\it irreducible} admissible cubics,
 are  the determinant  hypersurfaces   $ \cV_d = \{d = 1\} \cap \cV$  of  the  special Vinberg cones $\cV$.   Several  properties of such determinant hypersurfaces   and of their analogues   in   dual  spaces  have  relevant  implications in supergravity. In \cite{AMS} the theory of special Vinberg cones and their duals has been applied to compute the Bertotti-Robinson mass,  and therefore -- by  the Bekenstein-Hawking  formula --  the  entropy of static extremal BPS black holes (see \S \ref{sect44} for a detailed definition) in ungauged $N=2$, $D = 4$ supergravity with an arbitrary homogeneous vector multiplets' scalar manifold.
The theory of homogeneous convex cones has many other applications, for
example, in information geometry, multivariate statistics, Frobenius
manifolds, convex programming, etc.. For instance, for what concerns  the differential geometry of the convex (not necessarily
homogeneous) cones, Vinberg showed that the points of $\mathcal{V%
}$ define a probability measure on the adjoint cone $\mathcal{V}^{\ast }$, and
determined a canonical Riemannian Hessian metric $g$ on $\mathcal{V}$.   This means  that  the  homogeneous convex  cones  represent  a distinguished class of statistical Hessian manifolds, which in information geometry is  called   {\it exponential family}. For such Hessian manifolds, the above canonical  metrics  coincide with   the  {\it Fisher
metrics}  in statistics \cite{Am,Bar}.
\par
\smallskip
In this paper, we present a revision of Vinberg theory of homogeneous  convex cones and of their adjoint cones, which is based on the notion of {\it $\Nil$-algebras}, a particular class a nilpotent generalised matrix algebras introduced by Vinberg in \cite{Vi} (there, they  are  named  ``$N$-algebras") and
from which all Hermitian matrix algebras  corresponding to the Vinberg cones  can be recovered.\par
\smallskip In order to clarify  our approach,  let us  first briefly  recall the notion of homogeneous convex cone and of its adjoint one.
 Let $V = \bK^n$ be a vector space over $\bK = \bR$ or $\bC$ and denote by $\bK^*_o = \bR^+, \bC^*$ the connected component of the identity of
 the multiplicative group $\bK^*$.
 A {\it cone} $\cV \subset V$ is an open $\bK^*_o$-invariant domain, which contains no one-dimensional subspace.
It  is  called {\it homogeneous}  if  the linear  automorphism group
$$\Aut(\cV) = \left\{ \ A \in \GL(V)\ ,\ A(\cV) = \cV\ \right\}$$
   acts  transitively  on   $\cV$. It is called {\it of Vinberg type} if
    there  exists  a   solvable  subgroup $G \subset
    \Aut(\cV)  $
    which acts  transitively on  $\cV$ with a finite stabiliser $G_{x_o}$ at a point $x_o \in \cV$. In other words, $\cV$ is of Vinberg type if it has the form
    $$\cV = G(x_o) = G/G_{x_o}\quad \text{with}\ G\ \text{solvable and}\ G_{x_o}\ \text{finite}\ .$$
    Vinberg proved that {\it any homogeneous real convex cone $\cV$ is of this type} \cite{Vi}. Now,
given a    real vector space $V = \bR^n$, the {\it adjoint cone} of a cone $\cV \subset V$ is the cone in $V^* = \Hom(V, \bR)$ defined by
 $$ \cV^* = \{ \xi \in V^*,\,   \xi(x) >0\, \ \text{for all}\  x \in \cV  \}\ .$$
 It is known that if $\cV \subset V $ is convex, then  also  the adjoint cone $\cV^*$ is  convex and that a  convex cone $\cV$ is  homogeneous if and only if  its adjoint cone $\cV^*$ is homogeneous.
A homogeneous real convex  cone  $\cV \subset V = \bR^n$  is  called  {\it self-adjoint} (or {\it self-dual}) if there  exists  a Euclidean inner product  $\langle \cdot, \cdot \rangle$  such that   $ \cV^*  =\langle \cV, \cdot \rangle$.
\par
\smallskip
Vinberg  proved that any  homogeneous convex cone is  the  cone of positive definite matrices  in  some  algebra $\cH$ of generalised Hermitian matrices.
In \cite{Vi},  such algebra $\cH$  is introduced as a  subspace of a larger algebra of generalised matrices,  called {\it $T$-algebra}, which is in turn completely determined by its associative subalgebra of nilpotent upper triangular matrices,  called {\it $\Nil$-algebra}. In this paper, we re-formulate Vinberg theory of homogeneous cones  and their adjoint cones  in terms of    $\Nil$-algebras.
As a corollary, we get  that  the  pairs formed by a homogeneous cone and its  adjoint cone  are in bijection with the pairs of a $\Nil$-algebra $N$ and its  dual $\Nil$-algebra $N^{t^\prime}$, obtained from $N$ applying  the anti-transposition
(= reflection with respect to its anti-diagonal) to all  matrices. This implies that a homogeneous  cone is self-adjoint if and only if it is determined by a  $\Nil$-algebra $N$ that  is  isomorphic to its dual $N^{t^\prime}$.\par
\smallskip
We consider  the unimodular subgroup $G_0$ and the unipotent radical  $G'$  of the solvable Lie group $G$  acting simply transitively on a rank 2   and  special rank 3 Vinberg cone $\cV = G(x_o)$
and  determine the fields of the $G_0$-invariant  and the $G'$-invariant  rational functions on   $\cV$. As an application, we   determine  the $G_0$- and the  $G'$- invariant admissible cubics.   These give  one-parameter families  of cohomogeneity one and two  special real  manifolds corresponding to rank 2 and  special rank 3   Vinberg cones. In this way,   not only we recover  the well-known examples of homogeneous scalar manifolds determined by irreducible admissible cubics, but  we also obtain
\begin{itemize}[leftmargin = 20 pt]
\item[--]  a one-parameter family  of cohomogeneity one  scalar manifolds in  each Vinberg cone of  rank $2$;
\item[--] a one-parameter family of cohomogeneity two scalar manifolds  in  each  special  Vinberg cone of rank $3$.
\end{itemize}
\par
\medskip
The paper is structured as follows. In \S \ref{sect2} we give the definition of $\Nil$-algebra and we introduce    related objects that are needed for   the construction of the homogeneous  Vinberg cones. In \S 3, we describe the $G_0$- and $G'$-invariant rational functions on special Vinberg cones. In \S \ref{sect4} we describe the  $\Nil$-algebras that correspond  to  self-adjoint homogeneous  convex  cones (which include all rank 2  homogeneous convex  cones) and those associated with the  special  Vinberg cones of rank  3. The advertised description of the admissible cubics, corresponding to scalar manifolds of cohomogeneity one or two,  is given in \S\ref{sect5}.
\par
\medskip
\noindent{\it Notation}. In the following, for  any $m \times m$ matrix  $X = (x_{ij})$, $1 \leq i, j \leq m$, we denote by $t(X)$ and $t^\prime(X)$ the {\it transposed} and {\it anti-transposed} matrix of $X$, that is the matrix which is obtained from $X$ by reflection with respect to the diagonal and the anti-diagonal, respectively:
\begin{align*} &X = (x_{ij}) \xmapsto{\phantom{AAAAAAAAAA}} t(X) \= (x_{ji})\ &(\text{transposition})\\
 & X = (x_{ij}) \xmapsto{\phantom{AAAAAAA}} t^\prime(X) \= (x_{m+1 -j\, m - 1 + i})\ &(\text{anti-transposition})
 \end{align*}
 \par
\noindent{\it Acknowledgements.} The work of AM is supported by a ``Maria Zambrano'' distinguished researcher fellowship, financed by the European Union within the NextGeneration EU program. The authors are grateful to the referee for her/his useful comments and for  proposing  the interesting papers \cite{Bar,Mar}.
\medskip
\section{$\Nil$-algebras and  associated   Vinberg cones}\label{sect2}

\subsection{$\Nil$-algebras and  associated dual $\Nil$-algebras}
A {\it $\Nil$-algebra of rank $m$} over  $\bK = \bR$, $\bC$  is an  associative algebra $\cN$ of upper triangular  $m \times m$ matrices
\beq  \label{V} N = \begin{pmatrix}
0 & x_{12} & x_{13} & \ldots & \ldots & x_{1\, m-1} & x_{1\,m}\\
0 &  0 & x_{23} & \ldots & & x_{2\, m-1} & x_{2\,m}\\
0 &  0  & 0  & \ldots & & x_{2\, m-1} & x_{3\,m}\\
\vdots & \vdots & \ddots & \ddots &  \ddots &\vdots & \vdots\\
0 & 0& 0 & \ldots & \ldots &0 & x_{m-1\,m}\\
0 & 0& 0 & \ldots &\ldots &  0 & 0
\end{pmatrix} \ .\eeq
where  the entries $x_{ij}$ belong to  metric vector spaces $\cN_{ij}$, $i < j$, that is vector spaces   with  non-degenerate inner products $ \langle \cdot, \cdot \rangle_{\cN_{ij}}$. As a vector space,  $\cN$ is isomorphic to $\cN \simeq \sum_{i< j} \cN_{ij}$ and it is  equipped with the   inner product $\langle \cdot, \cdot \rangle = \sum_{i < j} \langle \cdot, \cdot \rangle_{\cN_{ij}}$.
The   matrix multiplication  in $\cN$ is defined by a system of  {\it isometric maps}, i.e.
  bilinear maps
 \beq \label{first} \bdot: \cN_{ij} \times \cN_{jk} \to \cN_{ik}\ ,\  (x_{ij}, x_{jk} ) \longmapsto x_{ij} \bdot x_{jk}\ ,\qquad i < j < k\ ,\eeq
 such that  \beq  \langle x_{ij} \bdot x_{jk}, x_{ij} \bdot x_{jk}\rangle {=} \langle x_{ij}, x_{ij} \rangle \langle x_{jk}, x_{jk} \rangle\ ,\eeq
 which  satisfy the following additional axiom:
 \begin{itemize}[leftmargin = 10pt]
 \item[] {\it   for  $x_{ik} {\in} \cN_{ik}$,  $x_{jk} {\in} \cN_{jk}$ with $i < j$}
\beq \langle x_{ik}, \cN\bdot x_{j k}\rangle {=}\{0\}\qquad \Longrightarrow\qquad \langle \cN \bdot x_{ik}, \cN \bdot x_{jk} \rangle{=} \{0\}\ .\eeq
\end{itemize}
Two $\Nil$-algebras $(\cN, \bdot)$, $ (\cN', \bdot^\prime)$ of  rank $m$ are  called  {\it isomorphic} if there is an algebra isomorphism $\f: \cN \to \cN'$ which preserves the inner products and  for which there is    a permutation $\s$ of $\{1, \ldots, m\}$  such that each subspace $\cN_{ij} $ is mapped into the subspace $\cN'_{\s(i)\ \s(j)} \subset \cN'$ and with the property
$$i < j\ \ \and\ \ \s(i) > \s(j)\qquad \Longrightarrow \qquad \cN_{ij} =  \cN'_{\s(i) \s(j)}=  \{0\}\ .$$
If $\bK = \bR$, a $\Nil$-algebra $\cN$ is called  {\it pseudo-Euclidean of signature $(p,q)$} (resp. {\it Euclidean}) if
its inner product $\langle \cdot, \cdot \rangle$ has signature $(p, q)$ (resp. is Euclidean). \par
\smallskip
For any  $\Nil$-algebra  $\cN$  we   associate  an algebra with inner product,  the {\it dual algebra $\cN^{t^\prime}$},  defined as follows.
Let
$\cN^{t^\prime}$ be the vector space of the matrices of the form
\beq  \label{V*} N^{t^\prime} = \left(\smallmatrix
0 & x^{t^\prime}_{12} & x^{t^\prime}_{13} & \ldots & \ldots & x^{t^\prime}_{1\, m-1} & x^{t^\prime}_{1\,m}\\
0 &  0 & x^{t^\prime}_{23} & \ldots & & x^{t^\prime}_{2\, m-1} & x^{t^\prime}_{2\,m}\\
0 &  0  & 0  & \ldots & & x^{t^\prime}_{2\, m-1} & x^{t^\prime}_{3\,m}\\
\vdots & \vdots & \ddots & \ddots &  \ddots &\vdots & \vdots\\
0 & 0& 0 & \ldots & \ldots &0 & x^{t^\prime}_{m-1\,m}\\
0 & 0& 0 & \ldots &\ldots &  0 & 0
\endsmallmatrix\right)\ .\eeq
 with entries  in the  vector   spaces $\cN^{t^\prime}_{ij} \= \cN_{m+1-j\, m+1-i}$,  equipped  with the
  inner product $\langle \cdot, \cdot \rangle = \sum_{i < j} \langle \cdot, \cdot \rangle_{\cN^{t^\prime}_{ij} \= \cN_{m+1-j\, m+1-i}}$.
 The   matrix  multiplication  $\bdot$ between elements of    $\cN^{t^\prime} $ is  given by the  family of isometric maps
\begin{multline*} \left(x^{t^\prime}_{ij} , x^{t^\prime}_{jk} \right)= \big(x_{(m+1 - j)\, (m+1 -i)} ,   x_{(m+1 - k)\, (m+1 -j)} \big) \xmapsto{\phantom{aaaaa}}  \\
\qquad \xmapsto{\phantom{aaaaa}}  x^{t^\prime}_{ij} \bdot  x^{t^\prime}_{jk}  \= x_{(m+1 - k)\, (m+1 -j)}\bdot x_{(m+1 - j)\, (m+1 -i)} \qquad
 \text{for any}\ i<j<k \end{multline*}
\smallskip
The following holds.
 \begin{lem} \hfill\par
 \begin{itemize}[leftmargin=20pt]
 \item[(1)] The   dual algebra   $(\cN^{t^\prime}, \bdot)$ is a $\Nil$-algebra;
 \item[(2)]  The anti-transposition operator
 \begin{multline}  \label{t'}  N = \left(\smallmatrix
0 & x_{12} & x_{13} & \ldots & \ldots & x_{1\, m-1} & x_{1\,m}\\
0 &  0 & x_{23} & \ldots & & x_{2\, m-1} & x_{2\,m}\\
0 &  0  & 0  & \ldots & & x_{2\, m-1} & x_{3\,m}\\
\vdots & \vdots & \ddots & \ddots &  \ddots &\vdots & \vdots\\
0 & 0& 0 & \ldots & \ldots &0 & x_{m-1\,m}\\
0 & 0& 0 & \ldots &\ldots &  0 & 0
 \endsmallmatrix \right)\longmapsto \\
 \longmapsto N^{t^\prime} = \left(\smallmatrix
0 & x_{m-1\, m} & \ldots & \ldots & \ldots & x_{2\, m} & x_{1\,m}\\
0 &  0 & x_{m-2\,m-1} & \ldots & & x_{2\, m-1} & x_{1\,m-1}\\
0 &  0  & 0  & \ldots & & x_{2\, m-2} & x_{1\,m-2}\\
\vdots & \vdots & \ddots & \ddots &  \ddots &\vdots & \vdots\\
0 & 0& 0 & \ldots & \ldots &0 & x_{1\,2}\\
0 & 0& 0 & \ldots &\ldots &  0 & 0
\endsmallmatrix\right)\end{multline}
 determines a linear (but, in general, not algebra) isomorphism between  $\cN$ and $\cN^{t^\prime}$.
\end{itemize}
 \end{lem}

\subsection{$\Solv$-algebras  and Vinberg groups}
Given a $\Nil$-algebra  $\cN$   of rank $m$  over $\bK$, the matrix multiplication  between elements in $\cN$ naturally extends to a matrix multiplication between  upper triangular  matrices of the form
\beq \label{V''} 
\begin{split}
& A = D + N
= \left(\smallmatrix
a_{11} & a_{12} & a_{13} & \ldots & \ldots & a_{1 m-1} & a_{1m}\\
0 &  a_{22} & a_{23} & \ldots & & a_{2 m-1} & a_{2m}\\
0 &  0  & a_{33} & \ldots & & a_{2 m-1} & a_{3m}\\
\vdots & \vdots & \ddots & \ddots &  \ddots &\vdots & \vdots\\
0 & 0& 0 & \ldots & \ldots & a_{m-1 m-1} & a_{m-1m}\\
0 & 0& 0 & \ldots &\ldots &  0 & a_{mm}
\endsmallmatrix\right)\\[10pt]
& \text{with} \ N \in \cN\ ,\  D {\=}\operatorname{diag}(a_{11}, \ldots, a_{mm}) \ \text{with}\ a_{ii} {\in} \bK\ ,
\end{split}
\eeq
 by means of  the natural maps
\beq\label{third} a_{ii}, a_{ik} \mapsto a_{ii}\bdot a_{ik}{\=}a_{ii} a_{ik} \ (= \text{usual product between scalar and  vector}).\eeq
The   extended   solvable algebra $\cT {=} \cT(\cN) {=} \bK^m {+} \cN$ of the matrices  \eqref{V''}
is called    {\it   $\Solv$-algebra associated with $\cN$}.
 \par
 \begin{definition}
 The {\it Vinberg group associated with $\cN$} is the connected component $G = G(\cN)$ of the identity $I_m = \operatorname{diag}(1, \ldots, 1)$
 of the solvable  Lie group of the invertible elements in $\cT = \cT(\cN)$.
 \end{definition}
  Note that a matrix  $A \in \cT$ is invertible if and only if its diagonal elements $a_{ii}$ are   non zero. Hence,
if $\bK = \bR$ (resp. $\bK = \bC$), the connected Lie group $G = G(\cN)$ consists of the matrices with positive diagonal elements (resp. of all invertible matrices) in $\cT$. \par
\smallskip
The Lie algebra $\gg = \gg(\cN)$ of $G(\cN)$ is given by the vector space  $\cT = \cT(\cN)$
equipped with the commutator
$$[A, A'] = A \bdot A' - A' \bdot A\ .$$
Note that the   Vinberg groups  $G(\cN)$, $G(\cN^{t^\prime})$, associated with a $\Nil$-algebra  $\cN$ and its dual algebra $\cN^{t^\prime}$, respectively,   are isomorphic  if and only if
$\cN, \cN^{t^\prime}$ are isomorphic $\Nil$-algebras.
\par
\medskip
\subsection{The $T$-algebra associated with  a $\Nil$-algebra}
Given a $\Nil$-algebra $\cN$ of rank $m$, we denote by $ \bT(\cN)$   the
vector space of all $m \times m$ matrices of the form
\beq M = \left(\smallmatrix
v_{1\,1} & v_{1\,2} & v_{1\,3} & \ldots & \ldots & v_{1\,m-1} & v_{1\,m}\\
v_{2\,1} &  v_{2\,2} & v_{2\,3} & \ldots & & v_{2\,m-1} & v_{2\,m}\\
v_{3\,1} &  v_{3\,2}  & v_{3\,3} & \ldots & & v_{2\,m-1} & v_{3\,m}\\
\vdots & \vdots & \ddots & \ddots &  \ddots &\vdots & \vdots\\
v_{m-1\,1 } & v_{m-1\,2 }& v_{m-1\,3} & \ldots & \ldots & v_{m-1\,m-1} & v_{m-1\,m}\\
v_{m\,1 } & v_{m\,2 }& v_{m\,3 } & \ldots &\ldots &  0 & v_{m\,m}
\endsmallmatrix\right)\ ,
\eeq
whose  entries $v_{ij}$ are elements of  the vector spaces
$$v_{ij} \in \left\{\begin{array}{ll} \cN_{ij}& \text{if} \ i < j\ ,\\[10pt]
\bK& \text{if}\ i = j\ ,\\[10pt]
\cN_{ji}^* = \Hom(\cN_{ij}, \bK) & \text{if}\ i > j\ .
\end{array} \right. $$
The space   $\bT(\cN)$ has   a canonical structure of {\it non-associative algebra over $\bK$} which  extends the  algebra structures  of  its subspaces  $\cN$ and  $\cT(\cN)$ \cite{Vi}. Such a structure
 is given by the unique system of bilinear maps $v_{ij} \times v_{jk} \mapsto v_{ik}$ for any $i,j,k$, which  coincide with the maps \eqref{first}, \eqref{third}  if $i \leq j \leq k$
 and satisfies  the following axioms:
  \beq
  \begin{split}
  &   (v_{ij}\bdot v_{jk})^\flat = v_{jk}^\flat \bdot v_{ij}^\flat,\\
  &  \langle v_{ij}^\flat \bdot v_{ik}, v_{jk}\rangle = \langle v_{ik}, v_{ij}\bdot v_{jk}\rangle,\\
  &\langle v_{ik}\bdot v_{jk}^\flat, v_{ij}\rangle = \langle v_{ik},v_{ij} \bdot v_{jk}\rangle ,
  \end{split}  \eeq
  where we use the notation $v_{ij}^\flat \= \langle v_{ij}, \cdot \rangle$. The algebra $\left(\bT(\cN), \bdot \right)$ is called  the  {\it Vinberg $T$-algebra} determined by   $\cN$. It is in general  {\it non-associative}
 but it   contains the   {\it associative} subalgebras  $\cN$ and $\cT(\cN)$.\par
\smallskip
By the results in \cite{Vi},  a  $\Nil$-algebra   $\cN$ is isomorphic to  its dual algebra  $\cN^{t^\prime}$   if and only if
the associated Vinberg $T$-algebras   $\bT(\cN)$ and $\bT(\cN^{t^\prime})$   are isomorphic.
 \par
\medskip
\subsection{Action of a Vinberg group on the space $\cH$ of  Hermitian matrices}
In the following, for  any $X  = (x_{ij}) \in \bT(\cN)$,  we  denote by $X^*$ the transposed matrix
$$X^* \= t(X^\flat) \in \bT(\cN)$$
where  $X^\flat = (x_{ij}^\flat)$ is the matrix with  co-vector  entries $x_{ij}^\flat$,  defined  by
$x_{ij}^\flat =\left\{\begin{array}{ll}  \langle x_{ij}, \cdot \rangle & \text{if}\ i \neq j\\[10pt]
x_{ii} & \text{if} \ i = j\end{array}\right.$.
Using this notation,  for any upper triangular matrix
$$A = D + N =  \left(\smallmatrix
a_{11} & a_{12} & a_{13} & \ldots & \ldots & a_{1 m-1} & a_{1m}\\
0 &  a_{22} & a_{23} & \ldots & & a_{2 m-1} & a_{2m}\\
0 &  0  & a_{33} & \ldots & & a_{2 m-1} & a_{3m}\\
\vdots & \vdots & \ddots & \ddots &  \ddots &\vdots & \vdots\\
0 & 0& 0 & \ldots & \ldots & a_{m-1 m-1} & a_{m-1m}\\
0 & 0& 0 & \ldots &\ldots &  0 & a_{mm}
\endsmallmatrix\right)  \in  \cT(\cN)\subset \bT(\cN)$$
we may consider the  {\it  associated Hermitian matrix}
$$H_A \= N^* + D + N = \left(\smallmatrix
a_{11} & a_{12} & a_{13} & \ldots & \ldots & a_{1 m-1} & a_{1m}\\
a_{12}^\flat &  a_{22} & a_{23} & \ldots & & a_{2 m-1} & a_{2m}\\
a_{13}^\flat &  a_{23}^\flat  & a_{33} & \ldots & & a_{2 m-1} & a_{3m}\\
\vdots & \vdots & \ddots & \ddots &  \ddots &\vdots & \vdots\\
a_{1 m-1}^\flat & a_{2 m-1}^\flat & a_{3 m-1}^\flat & \ldots & \ldots & a_{m-1\,m-1} & a_{m-1\,m}\\
a_{1 m}^\flat & a_{2 m}^\flat & a_{3 m}^\flat & \ldots &\ldots &  a^\flat_{m-1\,m} & a_{m\,m}
\endsmallmatrix\right) \in \bT(\cN)\ .$$
By construction,
$H^*_A = H_A$.
The vector  space of  such  matrices   is called  {\it  space $\cH = \Herm(\cN)$ of Hermitian matrices} \ associated with  $\cN$.  Notice  that the map $A \xmapsto{\phantom{aaaa}} H_A$ is a linear isomorphism  between $\cT(\cN)$ and $\Herm(\cN)$.
\par
\medskip
The next theorem collects  some of the crucial  results by Vinberg \cite{Vi} on  homogeneous cones  in  spaces of Hermitian matrices.
\begin{theo}  \hfill\par
    \begin{itemize}[leftmargin = 15pt]
\item[(1)]The map
$$L: \gg \times \Herm(\cN) \longrightarrow \Herm(\cN)\ ,\qquad L_A(X)\= A\bdot X + X \bdot A^*$$
is a faithful linear representation  of the Lie algebra $\gg = \gg(\cN) = \cT(\cN)$ of the Vinberg group $G = G(\cN)$ on $\Herm(\cN)$.
\item[(2)] The Lie algebra representation in (1) integrates to a  Lie group representation $\rho: G \to \GL(\cH)$ of the Vinberg group $G$ on $\Herm(\cN)$
\footnote{Such a $G$-action  is determined  by power series and  in general cannot be written in the standard form  $A (X) = A \bdot X \bdot A^*$}.
\item[(3)] The stabiliser $G_I$ of the $G$-action at the identity $I$ is a finite group (more precisely,
it is $ \{I\}$ if $\bK = \bR$ and it is $\{\operatorname{diag}(\pm 1, \ldots, \pm 1)\} \simeq \bZ_2^m$ if $\bK = \bC$) and the  orbit $ G(I) \simeq G/ G_I$ is
equal to the open cone  in the vector space $\cH = \Herm(\cN)$ given by
\beq \label{cone} \cV = \cV(\cN)\= \{ A \bdot A^*\ ,\  A \in G(\cN)\} \subset \cH\ .\eeq
\item[(4)]  If $\bK = \bR$ and  $\cN$ is  Euclidean,   the cone $\cV = \cV(\cN)$ defined in \eqref{cone} is  homogeneous and convex. Conversely, any homogeneous convex cone in a vector space   has the form $\cV = \cV(\cN) \subset \Herm(\cN)$ for some  Euclidean $\Nil$-algebra $\cN$.
\item[(5)] Two  convex cones  $\cV_1 = \cV(\cN_1)$ and  $\cV_2= \cV(\cN_2)$, associated   with  Euclidean  $\Nil$-algebras $\cN_1$, $\cN_2$, are linearly isomorphic if and only if $\cN_1$, $\cN_2$ are  isomorphic $\Nil$-algebras.
\end{itemize}
\end{theo}
The cone $\cV = \cV(\cN) \subset \cH$ defined in \eqref{cone} is called {\it Vinberg cone} determined by    $\cN$.
\par
\bigskip

\section{Invariant rational functions  of a   Vinberg  cone} \label{sect3}
\subsection{Matrix  and group coordinates of a Vinberg cone}
Let $\cN$ be a $\Nil$-algebra of rank $m$,   $G = G(\cN) \subset \cT(\cN)$ the associated Vinberg group and $\cV = G(I) \subset \Herm(\cN)$ the corresponding Vinberg cone. There are two natural multiplicative characters on the group $G$.
\begin{itemize}[leftmargin = 10pt]
\item[--] One of them is  the {\it $G$-determinant}, that   is the character that associates with any $A \in G$ the  product of its diagonal entries
$${\det}_G A = a_{11} \cdot \ldots \cdot a_{mm}\ .$$
\item[--] Another is the  {\it $\cH$-determinant}, that   is the determinant of the representation $\r$ of $G$ on  $\cH = \Herm(\cN)$
$${\det}_{\cH}: G \longrightarrow \bK\ ,\qquad {\det}_\cH(A) \= \det(\r_A)\ .$$
\end{itemize}
The map
$$\chi: \cV = G(I) \longrightarrow \bK\ ,\qquad X_A = A \bdot A^* \longmapsto \chi(X_A) \= \frac{1}{\det_\cH(A)}$$
is called the {\it Vinberg-Koszul characteristic function} of the cone $\cV = \cV(\cN)$. Note  that   it  is the density of a $G$-invariant measure on $\cV$. \par
\smallskip
Let us denote by $G' $ and  $ G_0$ the  normal subgroups of $G$ defined by
$$G' = \{\ A\in G\ , \ a_{ii} = 1\}\ ,\qquad G_0 = \{\ A\in G\ , \  {\det}_G A = 1\}\ .$$
We remark  that:
\begin{itemize}
\item[(a)] the (unimodular) subgroup $G_0$ is solvable and its Lie algebra is $\gg_0 = \cT(\cN)$;
\item[(b)]   $G'$ is the unipotent radical of $G$ and has   Lie algebra  $\gg' = \cN$.
\end{itemize}
\par
\bigskip
The entries  $x_{ij}$,  $i \leq j$,  of the elements $X = (x_{ij}) \in \cT(\cN)$ naturally determine coordinates on  $ \cH\simeq \cT(\cN)$ and on the  cone $\cV \subset \cH$,  which we call {\it matrix coordinates}.
Besides this,  on  the  cone $\cV \subset \cH$ there is  another   natural and very convenient set of coordinates, which we now introduce.
\par
\medskip
Recall that the matrix group $G$ is an open subset of the vector space $\cT = \cT(\cN)$ and that  the (vector) entries $a_{ij}, i \leq j$,   of the matrices  in $\cT$
 can be used to identify    $\cT$ with the vector space $\bK^{\dim \cT}$ and, consequently,     $G \subset \cT$ with an open subset of $\bK^{\dim \cT}$.
 Since the Vinberg  cone  $\cV = G(I)  $ is   in  bijection with  $G/G_I$ and  being $G_I= \bZ_2^m$  when $\bK = \bC$ and $G_I = \{I\}$ when $\bK = \bR$, the map
$$\psi_G: G \subset \bK^{\dim \cT} \longrightarrow  \cV \ ,\qquad A  \in G \subset \bK^{\dim \cT}  \overset{\psi_G} \longmapsto  X_A = A \bdot A^* ,$$
is a local diffeomorphism if $\bK = \bC$ and it is  a global diffeomorphism if $\bK = \bR$.  The composition
$$X \longmapsto A =   \xi_G(X) \longmapsto (a_{11}, a_{12} , \ldots, a_{ij}, \ldots )$$
of the  (local) inverse $\xi_G= \psi_G^{-1}$ with the  coordinates $a_{ij} = a_{ij}(A)$ of the matrices  $A \in G\subset \bK^{\dim \cT}$ can be considered as a set of coordinates $a_{ij} = a_{ij}(X)$ on $\cV$.  We  call them the  {\it  group coordinates} of the Vinberg cone.\par
\smallskip
The diagonal elements $a_{ii}(X)$, $1 \leq i \leq m$,  constitute a subset of  the group coordinates, which we call {\it diagonal group coordinates}.  Note that if $\bK = \bC$, the group coordinates of $\cV$ are {\it only locally} defined on $\cV$ and, in particular,   each diagonal group coordinate $a_{ii}(X)$, $1 \leq i \leq m$, is  defined only up to a sign.
\par
\medskip
\subsection{$G'$- and $G_0$- invariant rational functions of $\cV$}
A {\it $G'$- (resp. $G_0$-) invariant rational function   (resp. polynomial) of $\cV = G(I)$} is a rational function (polynomial) $q: \cV \to \bK$ of the matrix coordinates, which is invariant under any transformation $X \mapsto \rho_A(X)$,  $A \in G'$ (resp. $G_0$).  
\par
\medskip
Using group coordinates,  one can directly check
 that a rational function  $\f: \cV \to \bK$ is  $G'$-invariant if and only if it is a function only of  the  diagonal  group coordinates $a_{ii} = a_{ii}(X)$.  From this it follows that  a function $\f: \cV \to \bK$  is  $G_0$-invariant if and only if it has the form $\f(X) = \psi(\pi(X))$ where $\pi$ is the (locally defined) $\bK$-valued map given  by
 $$\pi(X) \=
  \prod_{i = 1}^m a_{ii}(X) $$
and $\psi$ is a rational function of one variable on $\bK$.
We recall that,  in case  $\bK = \bR$,   the diagonal group coordinates  $a_{ii}(X)$ (and hence also  the function  $\pi(X)$) are globally defined  on $\cV$.
On the other hand, if $\bK = \bC$,   the maps $a_{ii}(X)$ (and  hence  $\pi(X)$ too) are   locally defined with   values  determined  only up to a sign. This implies that    the   maps $a^2_{ii}(X)$ and $\pi(X)^2$ are  globally  defined in all cases.   \par
\medskip
Since for any $X_A = A A^* \in \cV$, the value  $\pi^2(X_A)$ is the square of the $G$-determinant of $A$, with a small abuse of language,   we  call  $\pi^2: \cV \to \bK$ the {\it squared $G$-determinant}.\par
\medskip
The next lemma is a crucial tool  to determine the invariant rational functions and  polynomials. It is a direct generalisation of the results of Vinberg in  \cite{Vi}. \par
\begin{lem}\label{lemma41}
There exist  $m$ homogeneous polynomials $p_i: \cV \to \bK$, $1  \leq i \leq m$,   of  degree $2^{m-i}$ in the matrix coordinates  of   $  \cV \subset \Herm(\cN)$
which allow to write the ({\rm globally defined}) $G'$-invariant functions  $a_{ii}^2(X)$, $1 \leq i \leq m$,  in terms of the matrix coordinates as
\beq \label{31} a_{ii}^2(X) = \frac{p_{i}(X)}{\prod_{s > i } p_s(X)} \eeq
In particular,  in  matrix coordinates, each function   $a_{ii}^2(X)$ is a homogeneous rational function of degree $1$. Moreover:
\begin{itemize}[leftmargin = 15pt]
\item[(i)] The  {\rm squared $G$-determinant} $\pi^2(X) =  \prod_{i = 1}^m a^2_{ii}(X)$ is a $G_0$-invariant  homogeneous rational function of degree $m$  in the matrix coordinates and  hence it extends  as a rational function over to the whole $\cH$ by the formula
\beq\label{ecco-1} \pi^2(X) = \prod_{i = 1}^m p^{-i + 2}_i(X) = \frac{p_1(X)}{\prod_{\ell \geq 3} p_\ell^{\ell -2}(X)}\  ;\eeq
\item[(ii)]
The {\rm Vinberg-Koszul characteristic function} $\chi: \cV \to \bK$ is  given in  matrix coordinates by {\rm (\cite[Ch. III, formula (33)]{Vi})}
\beq \chi(X) = \Pi_{i=1}^m p_i(X)^{n_i -n_{i-1}- \cdots - n_1 }  \ ,\qquad  \text{where}\qquad  n_i \=1 + \frac12 \sum_{s\neq i} \dim \cN_{is}  \ .\eeq
\end{itemize}
\end{lem}
\noindent Explicit expressions for  the  polynomials  $p_i(X)$ can be found  in \cite[\S III.3]{Vi}. From such expressions, one can  see that each  $p_i(X)$ is a polynomial of first order in the matrix coordinate $x_{ii}$ and is independent of the matrix coordinates $x_{jj}$, $j < i$. This implies that  the $m$-tuple of   polynomials $\big(p_1(X), p_2(X), \ldots, p_m(X)\big)$ is functionally independent  at all points. From  formula \eqref{ecco-1},  we   see that also the $m$-tuple $\left(\pi^2(X), p_2(X), \ldots, p_m(X)\right)$ is functionally independent at any point where $\pi$ is defined.
From Lemma \ref{lemma41} and these remarks we get:
\begin{prop} \label{propo} Let $p_i(X) $ be the polynomials of   Lemma \ref{lemma41}.Then:
\begin{itemize}[leftmargin = 15pt]
\item[(a)] Any   $G'$-invariant rational function $\f(X)$ on $\cV$ can be locally presented in the form
\begin{multline}
 \label{ecco} \f(X) = \psi\left(a_{11}(X), \ldots, a_{mm}(X) \right)= \\
 =  \psi\left( \sqrt{\frac{p_1(X)}{\prod_{s > 1} p_s(X)}}, \ldots,  \sqrt{\frac{p_{m-1}(X)}{\prod_{s > {m-1}} p_s(X)}},  \sqrt{p_{m}(X)}\right)\end{multline}
for appropriate locally defined square roots  and a locally defined real analytic function $\psi$; in case $\bK = \bR$, the maps $X \mapsto \sqrt{\frac{p_j(X)}{\prod_{s > j} p_s(X)}}$, $1 \leq j\leq m$,    are   globally defined on $\cV$ and $\psi$ is globally defined on the image of the map $$X \xmapsto{\phantom{aaaa}} \left( \sqrt{\frac{p_1(X)}{\prod_{s > 1} p_s(X)}}, \ldots,  \sqrt{\frac{p_{m-2}(X)}{ p_{m-1}(X) p_m(X)}}, \sqrt{\frac{p_{m-1}(X)}{p_m(X)}},  \sqrt{p_{m}(X)}\right)\ ;$$
In particular, a  polynomial $q(X)$ is $G'$-invariant  if and only if there is  a rational  function $Q(y_1, \ldots, y_m)$   that  allows to write $q(X)$ as
\begin{multline}
 \label{ecco-bis} q(X) = Q\left(a^2_{11}(X), \ldots, a^2_{mm}(X) \right)= \\
 =   Q\left( \frac{p_1(X)}{\prod_{s > 1} p_s(X)}, \ldots,  \frac{p_{m-2}(X)}{ p_{m-1}(X) p_m(X)},  \frac{p_{m-1}(X)}{ p_m(X)},  p_{m}(X)\right). \end{multline}
\item[(b)] The field   of rational functions  $\bK(\a_1, \ldots, \a_m) $,   generated by the ({\rm globally defined}) rational function  $\a_i(X) \= a_{ii}^2(X)$,
 is equal to
 $$\bK(\a_1,\a_2 \ldots, \a_m) = \bK(p_1, p_2, \ldots, p_m) = \bK(\pi^2, p_2, \ldots, p_m)$$
 and it
is therefore a subfield of the field of the  $G'$-invariant rational functions on $\cV$. In particular, a homogeneous polynomial is $G'$-invariant  if and only if it is  a polynomial in  $\bK(\pi^2, p_2, \ldots, p_m)$.
\item[(c)]   The field    $\bK(\pi^2)$,  generated by the squared $G$-determinant $\pi^2(X)$,  is  the  field of the $G_0$-invariant homogeneous rational functions  on $\cH = \Herm(\cN)$.
In particular, there exists a  $G_0$-invariant  homogeneous polynomial $q(X)$ of degree $m$ if and only if $\pi^2(X)$ is a polynomial and  $\operatorname{deg} \pi^2(X)$ divides $m$.  \end{itemize}
\end{prop}
\par
\bigskip

\subsection{Theory of duality} \label{theoryofduality}
Let $\bK = \bR$ and  $\cN$ be a Euclidean $\Nil$-algebra and denote by $G = G(\cN)$ and $\cV = \cV(\cN) = G(I)$ the corresponding Vinberg group and homogeneous Vinberg cone in
$\cH = \Herm(\cN)$, respectively.  We  denote by $G^* = \{ A^*, A \in G\} \subset \bT(\cN)$ the group of lower triangular matrices $A^* = t(A^\flat)$, $A \in G$. According to \cite{Vi, AMS},  the $G^*$-orbit of the identity is  the homogenous convex cone
$$ \cV' \= G^*(I) = \{ B B^*\ , \ B \in G^*\} = \{\ A^* A \ ,\ A \in G\}\ ,$$
which we call  {\it dual} {\it cone}. 
 The following proposition shows that the dual cone is metrically equivalent to the metric dual cone $\mathcal{V}^{\ast }:=g_{\mathcal{H}}\left( \mathcal{V}%
^{\prime }\right) $ and gives the  $\Nil$-algebra of which  it  is  the associated Vinberg cone.
\begin{prop} \label{duality}
Let  $g_{\cH}: \cH \to \cH^*$ denote the linear isomorphism determined by the Euclidean inner product  $g_{\cH} = \langle \cdot, \cdot \rangle$ of $\cH$.
\begin{itemize}[leftmargin = 15pt]
\item[(i)]  The   dual cone $\cV' =  G^*(I) $ is metrically equivalent to $\cV^*$ (that is, $g_\cH(\cV') = \cV^*$).
\item[(ii)] The anti-transposition  $t^\prime: \cH \to \cH^{t^\prime}$ maps isomorphically the dual cone $\cV' \subset \cH$ into the homogeneous cone $\cV(\cN^{t^\prime}) \subset \cH^{t^\prime}$
associated with $\cN^{t^\prime}$.
\item[(iii)] The cone $\cV$ is self-adjoint if and only if the $\Nil$-algebras $\cN$ and  $\cN^{t^\prime}$ are isomorphic (\cite{Vi}).
\end{itemize}
\end{prop}
\par
\medskip
\section{Examples of $\Nil$-algebras and  associated Vinberg cones}\label{sect4}
\subsection{\!$\Nil$-algebras of rank $2$ and  associated homogeneous cones}
The matrices of a   $\Nil$-algebra $\cN$ of rank $2$  have only one  non-trivial entry,  the entry $x_{12}$,  which belongs to the
 vector space $\cN_{12} = W = \bK^n$, equipped  with a non-degenerate inner product   $\langle \cdot ,\cdot \rangle $:
\beq \cN  = \cN_2(W) \= \left\{ \left( \begin{array}{cc} 0 & x_{12}\\ 0 & 0 \end{array}\right)\ ,\ \ x_{12} \in W\ \right\}\ .\eeq
The corresponding Vinberg group $G = G(\cN_2(W))$  consists of the group
\beq G =  \left\{ \left( \begin{array}{cc} a_{11} & a_{12}\\ 0 & a_{22} \end{array}\right)\  \text{where} \ \begin{array}{ll} a_{11}, a_{22} \neq 0& \text{if} \ \bK = \bC\\[10pt]
a_{11}, a_{22} > 0 & \text{if}\ \bK = \bR
\end{array}\ \ \text{and}\ \ a_{12} \in W \ \right\}\eeq
and acts on the  vector space $\cH_2(W)  = \Herm(\cN_2(W))$ of Hermitian matrices
\beq \cH_2(W) =  \left\{ X = \left( \begin{array}{cc} x_{11} & x_{12}\\ x^\flat_{12}  & x_{22} \end{array}\right)\  \text{where} \ x_{ii} \in \bK, \ x_{12} \in W \ \right\}\ .\eeq
This  vector space  has a natural structure of {\it  Jordan algebra},  given by the Jordan multiplication
$$X \circ Y \= \frac{1}{2} \left( X \bdot Y + Y \bdot X\right)\ .$$
If  $W$ is a Euclidean vector space, $\cH_2(W)$ is a Minkowski space and the Jordan algebra $(\cH_2(W), \circ)$  is
called {\it spin factor} \cite{Ba}.\par
\smallskip
The action of $G$ on $\cH_2(W)$ is given by the standard formula
$$\r_A(X) = A  \bdot X \bdot A^*$$
and the homogeneous  cone $\cV =  G(I) \subset \cH_2(W)$ is
\beq \begin{split}
 \cV = &  \left\{\  A \bdot A^* = \left(\begin{array}{cc} a_{11}^2 + \langle a_{12}, a_{12} \rangle & a_{22} a_{12} \\[10pt] a_{22} a_{12}^* & a_{22}^2 \end{array}\right) \ , \ a_{ii} \in \bK\ ,\ a_{12} \in W\ \right\} = \\
  = & \left\{\ X  \in W\   \text{with} \ \begin{array}{ll}  x_{11}x_{22} - \langle x_{12}, x_{12}\rangle \neq 0\ , \ x_{22} \neq 0& \text{if} \ \bK = \bC\\[10pt]
x_{11}x_{22} - \langle x_{12}, x_{12}\rangle > 0\ ,  \ x_{22} > 0 & \text{if}\ \bK = \bR \end{array}  \ \right\} \ .
\end{split}
\eeq
According to the explicit expressions in \cite{Vi}, the homogeneous polynomials $p_1(X)$ and $p_2(X)$  of  Lemma \ref{lemma41} are
\beq
\begin{split}
  & p_2(X)  =  a^2_{22} = x_{22}\ ,   \\
 & p_1(X) =  a^2_{11} a^2_{22}  =     x_{11} x_{22}-  |x_{12}|^2
\end{split}
\eeq
and the squared $G$-determinant  is  the quadratic polynomial
\beq
 \label{rank2} \pi^2(X) = p_1(X) =   x_{11} x_{22}-  |x_{12}|^2\ .
\eeq
If $\bK = \bR$ and $(W = \bR^n, \langle\cdot, \cdot \rangle)$ is  Euclidean,  then    $\cV = \cV(\cN_2(W)) $ is the cone of the future directed time-like vectors in the Minkowski space $\cH_2(W) \simeq \bR^{1, n+1}$ with Lorentz metric given by
$$\langle X, X \rangle = -  \pi^2(X) = - x_{11} x_{22} + | x_{12}|^2 \ .$$
In \cite{BH},  Baez and Huerta  gave a nice quantum mechanical interpretation of the homogeneous cone $\cV = \cV(\cN_2(W))$ in case the vector space   $W$ is one of the  division algebras $\bR, \bC, \bH$ or  $\bO$.

\subsection{Self-adjoint  homogeneous cones}
Let $\bD = \bR, \bC, \bH$ or $\bO$ be a division algebra. In the following, if  $\bD \neq \bO$,  we denote by
 $\cN_m(\bD)$ the $\Nil$-algebra of rank $m$
of the  upper triangular $m \times m$ matrices  with entries  $x_{ij}$, $i < j$,   in the algebra $ \bD$. If $\bD = \bO$, the $\Nil$-algebra  $\cN_m(\bO)$ is defined only  for $m = 3$.  One can check that:
\begin{itemize}[leftmargin = 15pt]
\item[--] the space  of  Hermitian matrices $\cH_m(\bD) \= \Herm(\cN_m(\bD))$ has a natural structure of Euclidean Jordan algebra given by the Jordan multiplication
$$X \circ Y \= \frac{1}{2} \left( X \bdot Y + Y \bdot X\right)\ $$
\item[--] the associated homogeneous cone $\cV_m(\bD) = \cV(\cN_m(\bD))$ is the self-adjoint cone of positive defined Hermitian $m \times m$ matrices over $\bD$.
\end{itemize}
M. K\"ocher and E.B. Vinberg, independently,  proved  the so-called  {\it K\"ocher-Vinberg Theorem} (see \cite{Koe, Vi0}) which establishes  a bijection  between the self-adjoint homogeneous convex cones and the Euclidean Jordan algebras.  This  result together with  the classification of Euclidean Jordan algebras by  Jordan, von Neumann and Wigner \cite{JNW} gives the following:
\begin{theo} \label{bigtheorem} There is a one-to-one correspondence between the self-adjoint homogeneous convex cones    and
 the Euclidean Jordan  algebras,   and any Euclidean Jordan algebra $(J, \circ)$ is a direct sum of  Jordan algebras of the form  $(\cH_2(W), \circ)$ or $(\cH_m(\bD), \circ)$. \par
 In particular, any
 indecomposable self-adjoint  homogeneous convex cone   $\cV$ is a symmetric space and it is either a  cone $\cV = \cV(W) \subset \cH_2(W)  $
 associated with  a $\Nil$-algebra of the form $\cN_2(W)$ or  a cone $\cV = \cV_m(\bD) \subset \cH_m(\bD)$ associated with  a  $\Nil$-algebra of the form $\cN_m(\bD)$.
 \end{theo}
\par
\medskip
\subsection{Special Vinberg cones   of  rank $3$}  \label{section43}
The matrices of a   $\Nil$-algebra $\cN$ of rank $3$  have only  {\it three}   non-trivial entries, namely  $x_{12} \in \cN_{12}$, $x_{13}\in \cN_{13}$ and $x_{23}\in \cN_{23}$.
The matrix multiplication between the matrices  in $\cN$ is  determined by the  isometric map
$$\bdot: \cN_{12} \times \cN_{23} \longrightarrow \cN_{13}\ ,\qquad (x_{12}, x_{23}) \longmapsto x_{12} \bdot x_{23}\ .$$
To simplify the notation,  we denote the spaces $\cN_{ij}$, $ i < j$,  by
$$V = \cN_{23}\ ,\qquad U = \cN_{12}\ ,\qquad W = \cN_{13}\ . $$
and   the  non-degenerate (= pseudo-Euclidean if $\bK = \bR$) inner products of $V$, $U$ and $W$ by $g_V = \langle\cdot, \cdot \rangle_{\cN_{12}}$,  $g_U = \langle\cdot, \cdot \rangle_{\cN_{23}}$   and  $g_W = \langle\cdot, \cdot \rangle_{\cN_{13}}$, respectively. We remark that:
\begin{itemize}[leftmargin = 12pt]
\item[--] If $\dim U = \dim W$, then the
  isometric   map
       $\bdot: V \times U \to W$ (i.e.  bilinear  map with
      \beq \label{isometricmap}  g_W(u\bdot v, u \bdot v)= g_V(v,v) \cdot g_U(u,u)\hskip 3 cm  \Big) \eeq
            defines a  structure   of
      $\bZ_2$ -graded  Clifford module with respect to   $\cC\ell(V, g_V)$ on the direct sum
              $$  S =S_0 + S_1 \=  U+ W\ ; $$
  \item[--]  Conversely,  if  $  S =S_0 + S_1$   is a $\bZ_2$ -graded  metric Clifford module over  $\cC\ell(V, g)$ and $g_{S_0}$ and $g_{S_1}$ are non-degenerate inner products on the spaces
  $S_0$, $S_1$ such that  \eqref{isometricmap} holds, then  the vector spaces $V$, $U \= S_0$ and $W \= S_1$ together with their inner products and the  Clifford multiplication $\mu: V \times S_0 \to S_1$
  determine a $\Nil$-algebra of rank $3$ with product $u \bdot v = \mu(v,u)$.
  \end{itemize}
      Due to this, the   description of all  $\Nil$-algebras of rank $3$ with  a prescribed metric vector space   $(V, g_V) = (\cN_{23}, \langle\cdot, \cdot \rangle_{\cN_{23}})$
      and with $\dim \cN_{12} = \dim \cN_{13}$  reduces to:
       \begin{itemize}[leftmargin = 15pt]
\item[(a)] the  classification      of $\bZ_2$-graded  Clifford
         $\cC\ell(V, g_V)$-modules
      \item[(b)]  the classification  of admissible  metrics in $S$ such that
      the  Clifford multiplication is  skew-symmetric.
      \end{itemize}
  The classifications (a) and (b) were obtained by Atiyah, Bott
         and  Shapiro  in \cite{ABS}   and  by   the first author and Cort\'es in \cite{AC}, respectively.\par
         \medskip
       These observations lead  to the  classification  of all homogeneous Vinberg cones $\cV(\cN) \subset \Herm(\cN)$ associated with
        $\Nil$-algebras $\cN$ of rank $3$ with
        $$\dim \cN_{12} = \dim \cN_{13}\ .$$
           In case $\bK = \bR$, such $\Nil$-algebras and their associated cones are called {\it special $\Nil$-algebras} and   {\it special Vinberg cones (of rank $3$)}, respectively (\cite{AC1}; see also \cite{AMS}) (\footnote{In our definition of a special Vinberg cone $\cV(\cN)$,   the   Nil-algebra  $\cN$  is   different  from the one   in  \cite{AC1, AMS}.   We     consider  
 matrices   $N = (N_{ij})$ in which $N_{12}$ is in $S_0$ (an even spinor)  and  $N_{23}$ is in   $V $ (a vector).  In  \cite{AC1, AMS},   the opposite   assumption is taken:
 $N_{12} \in V$ and   $N_{23} \in S_0$.  Only  our choice implies that   the squared $G$-determinant $d(X) = \pi^2(X)$ is a cubic polynomial,  while in the opposite case  it is a rational  function. Due to this,  some of  the  results  of  \cite{AC1,AMS}  should   be amended. A Corrigendum on this regard for \cite{AC1} and \cite{AMS} will be soon published.}).
       \par
       \smallskip
In detail, given  a  metric vector space  $(V, g)$  and an  associates   metric  $\bZ_2$-graded Clifford  $\cC\ell(V, g)$-module
$(S= S_0 + S_1, g_S)$  with skew-symmetric Clifford multiplication, we uniquely associate the  rank $3$  associative    $\Nil$-algebra
$\cN = \cN_3(V + S, g_V + g_S)$ given by the upper triangular matrices
\be \label{Triangulatmatrix-1}
  N =
   \begin{pmatrix}
0 & s_0 & s_1\\
0 & 0 & v\\
0 & 0      &  0
\end{pmatrix}\ ,
 \qquad v \in V, \  s_0 \in S_0, \  s_1\in S_{1}\ ,
\ee
with matrix multiplication given  by  $v \bdot s_0 \= \mu(v,s_0)$, where   $\mu$ stands for  the  Clifford multiplication.
The corresponding Vinberg group $G = G(V + S, g_V + g_S)$  consists of the group
\beq G =  \left\{ \left( \begin{array}{ccc} \a_1 & t_o & t_1\\ 0 & \a_2 &  w\\
0 & 0 & \a_3 \end{array}\right)\  \text{where} \ \begin{array}{ll} \a_i \neq 0& \text{if} \ \bK = \bC\\[10pt]
\a_i > 0 & \text{if}\ \bK = \bR
\end{array}\ \ \text{and}\ \ v\in V, t_i \in S_i\ \right\}\eeq
and acts on the  vector space
$$\cH_3(V{+}S) {=} \Herm(\cN_3(V{+}S, g_V {+} g_S))$$ of the Hermitian matrices
\beq \cH_3(V+S)=  \left\{ X = \left( \begin{array}{ccc} x_{1} &  s_0 & s_1\\ s_0^\flat & x_{2}  & v\\
s_1^\flat & v^\flat& x_{3} \end{array}\right)\  \text{with} \ x_{i} \in \bK, \ v \in V, s_i \in S_i \ \right\}\ .\eeq
According to the explicit expressions in \cite{Vi, AC1}, the   homogeneous polynomials $p_i(X)$  of  Lemma \ref{lemma41} are  given by
\begin{align}
 \nonumber & p_3(X)  =  a^2_{33} = x_3\ ,   \\
\nonumber & p_2(X) =  a^2_{33} a^2_{22}  =     x_3 x_2-  |v|^2 \ , \\
\nonumber &  p_1(X)  =a^2_{33} a^2_{22} a^4_{11} =x_3 d(X)\ ,
\end{align}
where $d(X) = a^2_{33} a^2_{22} a^2_{11} = \pi^2(X)$ is the cubic polynomial
\beq
 \label{Vinverses*} d(X) =    x_1x_2x_3  -  x_1  |s_0|^2 - x_2   |s_1|^2 -  x_3  | v |^2   + 2\langle \mu_v( s_0), s_1\rangle\ .
\eeq

The  homogeneous Vinberg cone  $\cV \subset \cH_3(V+S)$  is  given by Hermitian  matrices, whose matrix coordinates satisfy  the  inequalities:
\begin{itemize}[leftmargin = 10pt]
\item[--]   for $\bK = \bR$
\begin{align}
\nonumber & p_3(X) = x_3 > 0\ ,\qquad  p_2(X) = x_3 x_2-  |v|^2 > 0\ ,\\
 \label{positivity}
 &  d(X) = x_1x_2x_3  -  x_1  |  s_0|^2 - x_2   |s_1|^2 -  x_3  | v |^2   + 2\langle \mu_v(s_0), s_1\rangle> 0\ ;
 \end{align}
 \item[--] for  $\bK = \bC$
 \begin{align}
\nonumber & x_3 \neq  0\ ,\qquad  x_3 x_2-  |v|^2 \neq  0\ ,\\
 \label{positivity-1}
 &  x_1x_2x_3  -  x_1  |  s_0|^2 - x_2   |s_1|^2 -  x_3  | v |^2   + 2\langle \mu_v(s_0), s_1\rangle \neq 0\ .
 \end{align}
\end{itemize}
We recall   that  if  $g_V$ and $g_S = g_{S_0} + g_{S_1}$    are Euclidean,   then  the   $\Nil$-algebra $\cN = \cN_3(V + S, g_V + g_S)$  is Euclidean and the homogeneous Vinberg  cone
   $\cV \subset  \cH_3(V+S)$ is convex.
\par
\bigskip
\subsection{The  special  \textit{dual} Vinberg cones of rank $3$} \label{sect44}
By the results in \S \ref{theoryofduality},  the {\it  special {\rm dual} $\Nil$-algebras}, associated with   the special $\Nil$-algebras  over $\bK = \bR$,  determine the dual cones of
the special Vinberg cones. In detail, the  dual special $\Nil$-algebra  $\cN^{t'}$, associated with the $\Nil$-algebra determined by $(V, g_V)$ and $(S = S_0 + S_1, g_S)$, is the  matrix algebra
 of  upper triangular matrices
\be \label{Triangulatmatrix}
  N^{t^\prime} =
   \begin{pmatrix}
0 & v & s_1\\
0 & 0 & s_0\\
0 & 0      &  0
\end{pmatrix}\ ,
 \qquad v \in V,  \ s_0 \in S_0,  \ s_1\in S_{1}\ ,
\ee
with matrix multiplication given  by  $s_0 \bdot^\prime v \= \mu_v( s_0)$.
The corresponding  vector space of Hermitian matrices is
\beq \cH^{t^\prime}_3(V+S)=  \left\{ X = \left( \begin{array}{ccc} y_1 &  w & t_1\\ w^\flat & y_2  & t_0\\
t_1^\flat & t_0^\flat& y_3 \end{array}\right)\  \text{with} \ y_i \in \bK, \ w \in V, t_i \in S_i \ \right\}\eeq
and the  associated homogeneous Vinberg cone $\cV^{t^\prime}  \subset \cH^{t^\prime}_3(V+S)$ is  the open set
characterised by  the  inequalities :
\begin{align}
\nonumber & y_3 > 0\ ,\qquad  y_3 y_2-  |t_0|^2 > 0\ ,\\
 \label{positivity-2}
 &  d^{t^\prime}(Y) = y_1 y_2 y_3  -  y_3  | t_0|^2 - y_2   |t_1|^2 -  y_1 | w |^2   + 2\langle \mu_w(t_0), t_1\rangle
 +\\
 \nonumber & \phantom{aaaaaaaaaaaaaaaaaaaaaaaaaaaaaaaaaaaaaaa} +\frac{|t_0| ^2 |t_1|^2 - \langle t_1\bdot t_0^\flat, t_1 \bdot t_0^\flat\rangle}{y_3} > 0\ .
 \end{align}
 Moreover,  the   homogeneous polynomials    of  Lemma \ref{lemma41} are  in this case
\begin{align}
 \nonumber & p^{t^\prime}_3(X)  =  a^2_{33} = y_3\ ,   \\
\nonumber & p^{t^\prime}_2(X) =  a^2_{33} a^2_{22}  =     y_3 y_2-  |t_0|^2 \ , \\
\nonumber &  p^{t^\prime}_1(X)  =a^2_{33} a^2_{22} a^4_{11} =y_3 d^{t^\prime}(X)\ ,
\end{align}
where  $d^{t^\prime}(X) =  \pi^2(X)$ is the  squared $G$-determinant of $\cV^{t^\prime}$  given in \eqref{positivity-2}.
\par
\medskip
 Under the anti-transposition map $t^\prime: \cH^{t^\prime} \to \cH$, the Vinberg cone  $\cV^{t^\prime}  \subset  \cH^{t^\prime}_3(V+S)$ is mapped onto the
 dual cone $\cV' (\simeq \cV^*) \subset   \cH_3(V+S)$ of $\cV$,  characterised by the inequalities
 \begin{align}
\nonumber & x_1 > 0\ ,\qquad  x_1 x_2-  |s_0|^2 > 0\ ,\\
 \label{positivity*}
 &  d'(X) \= x_1 x_2 x_3  -  x_1  | s_0|^2 - x_2   |s_1|^2 -  x_3| v |^2   + 2\langle \mu_v(s_0), s_1\rangle + \\
\nonumber  & \phantom{aaaaaaaaaaaaaaaaaaaaaaaaaaaaaaaaaaaa} + \frac{|s_0| ^2 |s_1|^2 - \langle s_0\bdot s_1^\flat, s_0 \bdot s_1^\flat\rangle}{x_3}> 0\ .
 \end{align}
 Note that the  homogeneous rational function  $d'(X)$  of degree $3$ (associated  with the dual cone $\cV' = G^*(I) = (\simeq \cV^*) \subset \cH_3(V+S)$) differs from the cubic polynomial $d(X)$  of the Vinberg cone $\cV = G(I)$ by the term $\frac{|s_0| ^2 |s_1|^2 - \langle s_0\bdot s_1^\flat, s_0 \bdot s_1^\flat\rangle}{x^3}$. This shows that  for {\it any} special Vinberg cone,  the corresponding cubic polynomial  $d$ and the homogeneous rational function $d'$ of degree $3$  are  invariant under the actions on $\cH$ of the groups $G_0$ and
 and its dual group $G^*_0$, respectively.
This is a crucial property which,  in \cite{AMS},  allowed to compute  the Bertotti-Robinson mass -- and hence also the entropy -- of the BPS static,  spherically symmetric  and asymptotically flat BPS extremal black holes  in supergravity theory with homogeneous non-symmetric scalar manifolds.
 \par
 \smallskip
 From the above remarks   and from    Proposition \ref{duality}, Theorem \ref{bigtheorem}  and Hurwitz' classification of division algebras  we get
\begin{cor}\hfil\par
\begin{itemize}[leftmargin = 10pt]
\item[(1)]  For any special Vinberg cone $\cV$ (not necessarily self-adjoint), the cubic polynomial $d(X)$ and the homogeneous rational function $d'(X)$  of degree $3$  are  $G_0$- and $G^*_0$-invariant, respectively. 
\item[(2)] A special Vinberg cone $\cV \subset \cH_3(V+S)$  is self-adjoint if and only if  the spaces $V$, $S_0$ and $S_1$ are isomorphic Euclidean vector spaces. In this case $d = d'$.
 \end{itemize}
 \end{cor}
\par
\smallskip
\section{Construction of    special real manifolds with large isometry groups}\label{sect5}
\subsection {Admissible cubics}
Let $\cV = G(I)$ be a homogeneous Vinberg cone in the space of Hermitian matrices $\cH = \Herm(\cN)$ determined by a {\it real} $\Nil$-algebra $\cN$ of rank $m$.
Given a homogeneous  cubic polynomial $q: \cH \to \bR$,  we use the following notation:
\begin{itemize}[leftmargin = 10pt]
\item[--] $\cV_q$ is the hypersurface  in $\cV$  contained in  the level set  $\{ q = 1\}$:
$$\cV_q = \{X \in \cV\ : \ q(X) = 1\}\ ;$$
\item[--] $\Aut(q)$ is the group of linear transformations of $\cH$
which preserve $\cV$ and $q$:
$$ \Aut(q) = \{\ L \in \GL(\cH)\ :\ L(\cV) = \cV\ ,\ q(L(X)) = q(X)\ \text{for any}\ X\in \cH\ \}\ ;$$
\item[--] $g = g(q)$ is the  Hessian form given by $q$, i.e. the field  of symmetric bilinear forms on the hypersurface $\cV_q$ defined by
$$g_X(V, W) \= - \Hess(\log(q))|_X(V, W)\ ,\qquad X \in  \cV_q\ ,\ V, W \in T_X \cV_q\ .$$
\end{itemize}
The submanifold $\cV_q$ is called the {\it level hypersurface} of $q$ and $g = g(q)$ is  called  the {\it Hessian  form}   (or  {\it degenerate  metric}) of $\cV_q$. \par

\begin{definition} A  homogenous cubic polynomial   $q: \cH \to \bR$  is called {\it admissible} (resp. {\it locally admissible}) if the Hessian   form $g = g(q)$  is a Riemannian metric on $\cV_q$ (resp. on an open subset $\cU$ of $\cV_q$). In this case $(\cV_q, g)$ is called {\it special real manifold} \cite{WP1}.
\end{definition}
\par
\textbf{Remark}. One can easily see that the action of $\Aut(q)$ preserves the degenerate metric $g(q)$ on $\cV_q$.
\begin{lem}  \label{lemma52}
Let $q$ be a homogeneous cubic polynomial $q: \cH \to \bR$ and $H \subset \Aut(q)$  a subgroup of the  automorphism group of $q$. Then
$q$ is (locally) admissible if and only if the associated  Hessian form $g = g(q)$
 is positive definite at all  points of (an open subset of) a submanifold, which is transversal to the  $H$-orbits in $\cV_q$.
In particular:
\begin{itemize}[leftmargin = 20pt]
\item[(1)] If $G_0 \subset  \Aut(q)$, then $G_0$ acts transitively on $\cV_q$ and  $q$ is admissible if and only if $g = g(q)$  is positive definite at the identity matrix $I$.
\item[(2)] If $G' \subset \Aut(q)$, then  $G'$ acts on $\cV_q$  of cohomogeneity $m-1$  and $q$ is (locally) admissible  if and only if $g_X$  is positive definite  at all points $X$
of (an open subset of) the $(m-1)$-dimensional submanifold of $\cV_q$
$$\cD = \{\ X \in \cH\ : X = \operatorname{diag}(\lambda_1, \ldots, \lambda_m) \ \} \cap \cV_q\ .$$
\end{itemize}
\end{lem} 
\par
\medskip
\subsection{Admissible cubics on Vinberg cones of rank $2$}
Consider the real   $\Nil$-algebra of rank $2$
\beq \cN  = \cN_2(W) \= \left\{ \left( \begin{array}{cc} 0 & w\\ 0 & 0 \end{array}\right)\ ,\ \ w \in W\ \right\}\eeq
and let $\cV = G(I)$ be the homogeneous Vinberg cone in the space of  Hermitian matrices
 $$\cH_2(W) =   \left\{ X = \left( \begin{array}{cc} x_{1} & w \\ w^\flat  & x_{2} \end{array}\right)\  \text{where} \ x_{i} \in \bR, \  w \in W \ \right\} $$ associated with $\cN$.

\begin{theo}\label{theorem53} \hfill\par
\begin{itemize}[leftmargin = 15pt]
\item[(i)] There is no $G_0$-invariant cubic polynomial on $\cV$ and, in particular, no admissible cubic $q: \cV \to \bR$
which determines a $G_0$-homogeneous special real manifold.
\item[(ii)]Up to a  positive   scaling constant, the $G'$-invariant admissible cubics $q: \cV \to \bR$ are those having  the form
\beq \label{ecco-3} q(X) =  (x_{1} x_{2}-  |w|^2) x_2 + \ve x_2^3\ ,\qquad \ve \in \bR\ .\eeq
\end{itemize}
\end{theo}
\begin{pf} (i) Since  $\pi^2(X) $ is the irreducible polynomial of degree $2$
 \beq
 \label{rank2-bis} \pi^2(X) = p_1(X) =   x_{1} x_{2}-  |w|^2\ ,
\eeq
the claim follows immediately from Proposition \ref{propo} (c).
\ \\[10pt]
(ii) By Proposition \ref{propo} (a), any   $G'$-invariant polynomial   $q(X)$ on $\cV$  has  the form
\beq
 \label{ecco-1*} q(X)
 =  Q\left( x_2,  \frac{x_{1} x_{2}-  |w|^2}{x_2}\right) \eeq
for some   rational function  $Q(y_1, y_2)$. A function as in \eqref{ecco-1*}  is a homogeneous polynomial  of degree $3$ only if  the non-trivial monomials of the Taylor expansion of $Q$ at $0$ are only the   terms  of degree $3$, and thus only if  $Q$ itself is a homogeneous polynomial of degree $3$. This implies  that $q(X)$ is a $G'$-invariant  homogeneous cubic   if and only if  it has the form
\beq  \label{ecco-2} q(X)
 =   a x_2^3 + b x_2 \left(x_{1} x_{2}-  |w|^2\right) = x_2 \left(a x_2^2 +   b \left(x_{1} x_{2}-  |w|^2\right) \right)\eeq
 for some real coefficients $a$, $b$.  Since for $b = 0$ this  cubic   is associated with a symmetric bilinear form $g = g(q)$ on the level hypersurface $\cV_q = \{ x^3_2 = 1\}$ which is degenerate,  the cubic \eqref{ecco-2} is admissible only if $b \neq 0$. Hence, up to a scaling, $q(X)$  has   the form \eqref{ecco-3}.  By Lemma \ref{lemma52}, a cubic $q(X)$ is admissible if and only if, for any diagonal matrix $X = \operatorname{diag}(x_1, x_2) \in \cV_q$, the Hessian form    $g(q) =  - \Hess(\log(q))|_{T \cV_q}$
 is well defined and  positive definite   at   $X$.
          A direct computation shows that for any  $X = \operatorname{diag}(x_1, x_2) \in \cV_q$, the Hessian  $n \times n$ matrix
           $H_X= - \Hess(\log(q))_X$,  $n= \dim \cH$,  has the  form
\begin{equation}
- \Hess(\log(q))_X=\left(
\begin{array}{cc}
 \smallmatrix \frac{1}{(x_1+\varepsilon x_2)^2} & \frac{\ve}{(x_1+\varepsilon x_2)^2}  \\ \frac{\ve}{(x_1+\varepsilon x_2)^2}  & \frac{1}{(x_1+\varepsilon x_2)^2} \left(2\left(
\frac{x_1}{x_2}\right) ^{2}+4\varepsilon \frac{x_1}{x_2}%
+3\varepsilon ^{2} \right) \endsmallmatrix & 0 \\
  0 & \frac{2 I_{n-2}}{x_1 x_2+\varepsilon
x_2 ^{2}}%
\end{array}%
\right) \ .
\end{equation}
Therefore claim (ii) follows by observing  that  for any $\ve \in \bR$
\begin{itemize}[leftmargin = 20pt]
\item[(a)] $- \Hess(\log(q))_X$ (and its restriction to $g(q)$) is well defined at all points of
$$\{\operatorname{diag}(x_1, x_2)\}  \cap \cV_q = \{\operatorname{diag}(x_1, x_2)\ , \ x_2  > 0\ ,\ x_2( x_1 +  \ve x_2)= 1\}$$
\item[(b)] $- \Hess(\log(q))_X$  is  positive definite  at all points of $\{\operatorname{diag}(x_1, x_2)\}  \cap \cV_q$.
\end{itemize}
Indeed, by  Sylvester criterion,   (b)  can be checked by just computing   the principal second order minor
\begin{eqnarray*}
\frac{1}{(x_1+\varepsilon x_2)^4} \det\left(\begin{array}{cc}
1 & \varepsilon \\
\varepsilon & \left(2\left(
\frac{x_1}{x_2}\right) ^{2}+4\varepsilon \frac{x_1}{x_2}%
+3\varepsilon ^{2}\right)\end{array} \right) =\\
= \frac{2}{(x_1+\varepsilon x_2)^4} \left( \varepsilon +\frac{x_1}{x_2}\right) ^{2} = \frac{2}{x_2^2 (x_1+\varepsilon x_2)^2}\ ,
\end{eqnarray*}
which  is manifestly  positive on  $\{\operatorname{diag}(x_1, x_2), x_1, x_2 > 0,  x_2(x_1 + \ve x_2)  = 1\}$.
\end{pf}
\par
\medskip

\subsection{Admissible cubics on special Vinberg cones of rank $3$}
\par \medskip
Let $\cV = G(I)$ be a special  Vinberg cone in the space of  Hermitian matrices

\beq \cH_3(V+S)=  \left\{ X = \left( \begin{array}{ccc} x_{1} &  s_0 & s_1\\ s_0^\flat & x_{2}  & v\\
s_1^\flat & v^\flat& x_{3} \end{array}\right)\  \text{where} \ x_{i} \in \bR, \ v \in V, s_i \in S_i \ \right\}\ .\eeq
determined a special $\Nil$-algebra $\cN = \cN_3(V + S, g_V + g_S)$. \par

\begin{theo}\label{theorem54}\hfill\par
\begin{itemize}[leftmargin = 15pt]
\item[(i)] Up to a  scaling, there is only one  $G_0$-invariant cubic polynomial on $\cV$, namely the squared $G$-determinant \eqref{Vinverses*}. This cubic polynomial
is admissible if and only if the $\Nil$-algebra $\cN_3(V + S, g_V + g_S)$ is Euclidean.
\item[(ii)]
The $G'$-invariant  polynomial algebra is the   polynomial algebra $\bK[d, p_2, p_3]$. In particular any $G'$-invariant cubic is a linear combination
of the cubics
\beq \label{threecubics} d(X) \ ,\qquad p_2(X) p_3(X) =  x_3 \left(x_{2} x_{3}-  |v|^2\right)\ ,\qquad p^3_3(X) = x_3^3\eeq
and any  linear combinations of these  cubics is an admissible cubic only if, up to a positive scaling,  it has the form
\beq \label{form-3} q(X) =  d(X) +  \ve_1 x_3 \left(x_{2} x_{3}-  |v|^2\right) + \ve_2 x_3^3\ \  \text{with}\  \ve_1 \in \bR \ \text{and}\ \ve_2  \leq 0\ .\eeq
Moreover: \begin{itemize}
\item For any pair $(\ve_1,\ve_2=0)$ the corresponding cubic is admissible;
\item For the pairs $(\ve_1,\ve_2)$ with $\ve_2<0$, $\ve_1=\alpha|\ve_2|$, with $\alpha>0$, the corresponding cubics are locally admissible.
\end{itemize}
\end{itemize}
\end{theo}
\begin{pf}
(i)  We recall that  the squared $G$-determinant $\pi^2(X)$ is  the irreducible cubic polynomial    \eqref{Vinverses*}.  By  Proposition \ref{propo} (c),  up to a scaling,
$d(X) = \pi^2(X)$ is the unique  $G_0$-invariant homogeneous polynomial  of degree $3$.  The second claim  can be checked by  explicitly determine the  expression of the symmetric bilinear form $g|_I = - \Hess(\log(d))|_I$
in matrix coordinates.  \par
\smallskip
(ii) The same arguments for the claim (ii) in Theorem \ref{theorem53} show that  the  $G'$-invariant cubic polynomials $q(X)$ are exactly those  which are linear combinations of   the three cubics \eqref{threecubics}, namely
\beq  \label{ecco-31} q(X)
 =   a\, d(X)  + b\, p_2(X) p_3(X) + c\, p^3_3(X)\ ,\qquad a,b,c \in \bR\ .\eeq
One can directly check that  for $a = 0$, the corresponding cubic is associated with a degenerate metric $g = g(q)$.  Hence a cubic $q(X)$  is admissible only if,  up to a scaling, it has   the form
 \beq \label{form-3*} q(X) =  d(X) +  \ve_1 x_3 \left(x_{2} x_{3}-  |v|^2\right) + \ve_2 x_3^3\ ,\ \ve_1, \ve_2 \in \bR\ .\eeq
  By Lemma \ref{lemma52}, this cubic is admissible if and only if, for any matrix $X$ in the set
  \begin{multline} \label{transvers}  \{ \operatorname{diag}(x_1, x_2, x_3)\} \cap  \cV_q =\\
  =  \{\operatorname{diag}(x_1, x_2, x_3)\ ,\ x_i> 0\ , \ x_3\left(x_1 x_2 + \ve_1 x_2 x_3 + \ve_2 x_3^2\right)= 1\}\end{multline}
  the symmetric form   $H_X := - \Hess(\log(q))_X$
 is well defined and its restriction to $T_X \cV$  is positive definite. A direct computation shows that for any  $X $ in \eqref{transvers},  $H_X =- \Hess(\log(q))_X$ reads
 \newcommand{\bigzero}{\mbox{\normalfont\large 0}}
\begin{multline} \label{theH}
- \Hess(\log(q))_X = q^{-2}(X)\cdot \hskip 5 cm \\
\hskip 1 cm {\cdot} \left(
\begin{array}{ccc}
\smallmatrix
 x_{2}^{2} x_3^{2} & -\varepsilon_{2}
x_3^{4} &\cA\\
-\varepsilon _{2} x_3^{4} & \cB&\cC  \\
\cA &\cC & \cD \endsmallmatrix & \vline & \bigzero  \\[10pt]
\hline
\bigzero & \vline &  \smallmatrix 2q(X)x^{3}I_{\dim V} & 0 & 0 \\
 0 & 2q(X)\left( x_1+\varepsilon _{1}x_3\right)I_{\dim S/2} & 0 \\
 0 & 0  & 2q(X)x_2 I_{\dim S/2} \endsmallmatrix%
\end{array}
\right)  \end{multline}
where
\beq
\begin{split}
&\cA \= x_2 x_3
^{2}\left(\varepsilon _{1}x_2 +2\varepsilon_{2}x_3\right) \ ,\qquad
\cB \= x_3^2 \left( x_{1}+\varepsilon _{1}x_3\right)
^{2} \ ,\\
&\cC \= \varepsilon _{2} x_{3}
^3\left( 2x_{1}+\varepsilon _{1} x_{3}\right) \ , \\
&\cD \=  x_2^2 \left( x_1+\varepsilon _{1} x_3\right)
^{2}+ x^2_{3}\left( \varepsilon _{1}x_{2}+\varepsilon _{2} x_{3}\right) ^{2}+2\varepsilon _{2}x^3_{3}\left( \varepsilon _{1}x_{2}+\varepsilon _{2} x_3\right)
\end{split} \eeq
From this expression it is manifest that $g(q)_X=  H_X |_{T_X{\cV_q}}$ is positive definite only if the following inequality holds:
$$x_1+\varepsilon _{1}x_3 > 0\ .$$
Since we are assuming  that $X$ is in \eqref{transvers}, this condition is equivalent to
\beq \label{constr1} 1 - \ve_2 x_3^3 > 0\ .\eeq
Now,   for any fixed value of $\ve_2 > 0$, there is  at least one  point $X_o$ in \eqref{transvers} with sufficiently large coordinate $x_3$ so that
\eqref{constr1} is not satisfied. We may therefore  conclude that if $\ve_2 > 0$ the corresponding cubic is not admissible. Moreover, in the case $\ve_2 = 0$ the Hessian  $H_X =- \Hess(\log(q))_X$
has the form
\beq
 H_X =
 \left(
\begin{array}{ccc}
\smallmatrix
 x_{2}^{2} x_3^{2} & 0 & \ve_1 x^2_2 x^2_3
 \\
0 &  x_3^2 \left( x_{1}+\varepsilon _{1}x_3\right)
^{2} & 0  \\
 \ve_1 x^2_2 x^2_3 &0 &   x^2_2( x_1 +  \varepsilon _{1} x_3)^2 +   \ve^2_1 x_2^2 x_3^2
\endsmallmatrix&\vline & \bigzero  \\[10pt]
\hline
\bigzero & \vline & \ast
\end{array}
\right)\ . \eeq
  A direct computation based on Sylvester criterion shows that it is  positive definite at each point of \eqref{transvers} for any choice of $\ve_1$.

Finally, in the cases when  $\ve_2<0$ and $\ve_1=\alpha|\ve_2|$, with $\alpha>0$, a tedious but straightforward computation shows the existence of a point $X_o$ in the set \eqref{transvers} at which $H|_{X_o}$ is positive definite. This implies that there exists a neighbourhood of $X_o$ on which $H$ is positive definite, meaning that the corresponding cubic is locally admissible.
\end{pf}

\par
\medskip

\bibliographystyle{amsplain}

\end{document}